\documentclass[a4paper,11pt]{article}
\usepackage{jcappub} 
\usepackage{lineno}

\title{\boldmath Curvature and Isocurvature Perturbations in multi-field Gauss-Bonnet inflation}

\author{Seoktae Koh}
\affiliation{Department of Science Education, Jeju National University,\\
Jeju, 63243, South Korea}

\emailAdd{kundol.koh@jejunu.ac.kr}

\abstract{We study cosmological perturbations in multi-field inflation in which the scalar fields couple to the Gauss-Bonnet terms through a general coupling function $f(\phi^a)$. We derive the complete quadratic action for the field perturbations in spatially flat gauge and find that, besides corrections to the kinetic, gradient and mass matrices, the Gauss-Bonnet coupling induces an antisymmetric velocity coupling that vanishes in Einstein gravity. Decomposing the field perturbations into curvature and isocurvature modes,  we show that the effective mass of the curvature perturbation and the non-derivative mixing mass vanish identically, $\mathcal{M}_{\mathcal{R}}^2 = \mathcal{M}_{\rm mix}^2 =0$ which are justified by the Weinberg's adiabatic mode. We then compute the curvature power spectrum in two limits. First, when the isocurvature mode is heavy, it can be integrated over all scales. This yields an effective single-field theory of the curvature perturbation with a modified sound speed. Second, when the gradient of the coupling function is aligned with the background trajectory ($f_N =0$), the derivative mixings disappear and the superhorizon transfer of isocurvature into curvature perturbations is governed  by the turn rate, which is modified by the Gauss-Bonnet corrections. We also present the tensor power spectrum and the tensor-to-scalar ratio at linear order in the coupling. 
}

\begin{document}
\maketitle
\flushbottom

\section{Introduction}
\label{sec:intro}

Despite the remarkable success of single-field slow-roll inflation in describing current observations, a generic ultraviolet completion of inflation contains multiple scalar degrees of freedom. In particular, string compactifications and supergravity theories typically contain moduli, axions, and other scalar fields, so that obtaining an effective single-field description requires the remaining fields to be sufficiently stabilized or decoupled \cite{Douglas:2006es}. This motivates studying inflation with multiple dynamical scalar fields, where the geometry of the scalar field space and the evolution of adiabatic and isocurvature perturbations play an essential role. Multifield dynamics can lead to effects absent from the simplest single-field description, including the sourcing of curvature perturbations by isocurvature modes \cite{Gordon:2000hv, GrootNibbelink:2001qt, Lalak:2007vi}, destabilization of the inflationary trajectory due to field-space curvature \cite{Renaux-Petel:2015mga}, transient enhancement of the curvature power spectrum associated with sharp turns of the inflationary trajectory \cite{Fumagalli:2019noh}, and characteristic signatures of heavy fields, including a reduced effective speed of sound and oscillatory features \cite{Achucarro:2010da, Cespedes:2012hu}, as well as distinctive non-Gaussian signatures associated with quasi-single-field inflation and cosmological collider physics \cite{Chen:2009zp, Arkani-Hamed:2015bza}.

At the same time, higher-curvature corrections are naturally expected in an effective description descending from a ultraviolet completion of gravity. Among the possible higher-curvature operators, the Gauss-Bonnet (GB) invariant is particularly interesting because, when coupled to a dynamical scalar sector, it modifies the inflationary background and perturbation dynamics while maintaining second-order equations of motion. The cosmology of Einstein-scalar-Gauss-Bonnet gravity has been studied extensively in the single-field context, including primordial perturbations \cite{Satoh:2008ck}, slow-roll inflation and its observational constraints \cite{Guo:2010jr,Jiang:2013gza,Koh:2014bka} and primordial gravitational wave \cite{Bernardo:2025lie}. 

This extension is particularly interesting because the multifield structure introduces new channels through which the GB sector can affect inflationary dynamics. In a multifield theory, the coupling $f(\phi^a)$ is a scalar function on the curved field space and its covariant gradient can have both adiabatic and isocurvature components. Consequently, the GB interaction can couple directly to the entropic sector and modify the effective mass, mixing, and evolution of curvature and isocurvature perturbations. Moreover, the GB-induced modification is intertwined with the field-space geometry and the turning of the inflationary trajectory. These effects have no direct analogue in the single-field theory, where there is no independent isocurvature direction.

To date, however, these two directions have remained largely disconnected: the perturbation theory of Gauss-Bonnet inflation has been studied predominantly in a single field setting. In this paper, we take a first step toward combining them and study curvature and isocurvature perturbations when several scalar fields, living on a curved field space with metric $G_{ab}$, couple to the Gauss-Bonnet term through an arbitrary function $f(\phi^a)$. We derive the quadratic action for scalar and tensor perturbations and identify the modifications to the coupled curvature-isocurvature system induced by the GB interaction. We further investigate the resulting power spectra in the heavy-isocurvature regime and in the $f_N=0$ and weak-coupling limits.

The paper is organized as follows. In section \ref{sect:multi-inflation} we introduce the model and derive the background equations. In section \ref{sect:quad-action}) we obtain the quadratic action for the scalar and tensor perturbations, perform the curvature-isocurvature decomposition, and the consistency conditions implied by Weinberg's adiabatic mode. In section \ref{sect:spectra} we compute the power spectra in the heavy isocurvature mode limit and in the $f_N = 0$ and weak coupling limit. We summarize and discuss our results in section \ref{sect:summary}

\section{Multi-field inflation with a Gauss-Bonnet coupling}\label{sect:multi-inflation}

We consider multi-scalar fields $\phi^a$ coupled to the Gauss-Bonnet term
\begin{align}
	S = \int d^4x \sqrt{-g} \biggl[\frac{M_{p}^2}{2} R - \frac{1}{2} G_{ab} g^{\mu\nu} \partial_{\mu} \phi^a \partial_{\nu} \phi^b - V(\phi^a) + \lambda f(\phi^a) R_{GB}^2 \biggr],
    \label{multi-action}
\end{align}
where $M_{p}^2 = (8\pi G)^{-1}$ is the reduced Planck mass, $G_{ab}$ is the field space metric, the indices $a,\,b,\,\cdots$ label the fields, $f(\phi^a)$ is the coupling function, $\lambda$ is a dimensionless coupling parameter, and the Gauss-Bonnet invaraint is 
\begin{align}
    R^2_{GB} = 	R^2 - 4 R_{\mu\nu} R^{\mu\nu} + R_{\mu\nu\rho\sigma} R^{\mu\nu\rho\sigma}.
\end{align}

With the flat Friedmann-Lemaitre-Robertson-Walker (FLRW) metric
\begin{align}
    ds^2 = -dt^2 + a^2(t) \delta_{ij} dx^i dx^j,	
    \label{flrw_metric}
\end{align}
the background equations of motion are obtained by varying the action (\ref{multi-action}) with respect to $g_{\mu\nu}$ and $\phi^a$,
\begin{align}
    & 3 M_{p}^2 H^2  -  \frac{1}{2} \dot{\phi}^2 - V_0 + 24\lambda H^3 \dot{f} = 0,
 \label{friedmeq-multi}
\\
 & D_t \dot{\phi}_0^a + 3 H \dot{\phi}_0^a + \bar{G}^{ab} \partial_b V - 24\lambda H^2\bar{G}^{ab} ( H^2 + \dot{H} ) f_b= 0,
  \label{kgeq-multi}
\end{align}
where $\dot{\phi}^2 = \bar{G}_{ab} \dot{\phi}^a_0 \dot{\phi}^b_0,\,\, \dot{f} = f_a \dot{\phi}^a_0$ with $f_a \equiv \partial f/\partial \phi^a$, and the covariant time derivative along the trajectory is defined by \cite{GrootNibbelink:2001qt, Gong:2016qmq}
\begin{align}
	D_t \dot{\phi}_0^a = \ddot{\phi}_0^a + \Gamma^a_{bc} \dot{\phi}_0^b \dot{\phi}_0^c,
\end{align}
with $\Gamma^a_{bc}$ the Christoffel connection of the field-space metric $G_{ab}$. The overdot denotes the derivative with respect to cosmic time $t$, and the field-space indices are raised and lowered with the field-space metric $G_{ab}$. Bar quantities denote background quantities. Combining the time derivative of  eq. (\ref{friedmeq-multi}) with eq. (\ref{kgeq-multi}), one obtains the evolution for the Hubble parameter,
\begin{align}
    6 M_{p}^2 \dot{H} =&\,\, - 3 \dot{\phi}^2  + 24\lambda H^3 \dot{f} - 48\lambda H \dot{H} \dot{f} - 24\lambda H^2 \ddot{f}.
    \label{dH-multi}
\end{align}

Notice that with the FLRW  background (\ref{flrw_metric}), the Gauss-Bonnet invariant  evaluates to $\bar{R}_{GB}^2 = 24 H^2 (H^2 + \dot{H})$, so the Gauss-Bonnet contribution to the field equation (\ref{kgeq-multi}) is simply $\lambda \bar{G}^{ab} f_b \bar{R}_{GB}^2$ . The coupling therefore acts as an additional, background-dependent potential force, and the field-space gradient that drives the trajectory is that of the effective potential
\begin{align}
    V_a^{\rm eff} = V_a - 24\lambda H^2 (H^2 +\dot{H}) f_a,	
    \label{eff_pot}
\end{align}
which will be useful when we discuss the bending of the trajectory below.

\section{Quadratic action} \label{sect:quad-action}

\subsection{Scalar perturbations in the spatially flat gauge}
We adopt the Arnowitt-Deser-Misner (ADM) form of the metric to study the perturbations,
\begin{align}
	ds^2 = -N^2 dt^2 + \gamma_{ij} \left(N^i dt + dx^i\right) \left(N^j dt + dx^j \right),
 \label{admmetric}
\end{align}
in terms of which the action (\ref{multi-action}) becomes \cite{Julie:2020vov}
\begin{align}
 \mathcal{S} =&\,\, \int d^4x \sqrt{\gamma} \biggl[ \frac{M_{p}^2}{2}\biggl( N \tilde{R} - N^{-1} (E^2 - E^{ij} E_{ij}) \biggr) - \frac{1}{2} \gamma^{ij} G_{ab}\partial_i \phi^a \partial_j \phi^b + 2 N^{-1} G_{ab} E_{\phi}^a E_{\phi}^b 
    \nonumber \\
    &\,\,- N V(\phi)   + 2\lambda \biggl\{ - N \biggl( 2 \tilde{R}  \tilde{\nabla}^2 f(\phi) - 4 \tilde{R}^{ij} \tilde{\nabla}_i \tilde{\nabla}_j f(\phi) \biggr) + 2 N^{-1} \biggl( (E^2 - E^{ij} E_{ij}) \tilde{\nabla}^2 f(\phi) 
    \nonumber \\
    &\,\, + 2 E^{ij} E_{jk} \tilde{\nabla}^k \tilde{\nabla}_i f(\phi) - 2E E^{ij} \tilde{\nabla}_i \tilde{\nabla}_j f(\phi) \biggr) + 2 f_a(\phi) N^{-1} E^a_{\phi} (2 E \tilde{R} - 4 E^{ij} \tilde{R}_{ij}) \nonumber \\
    &\,\, + \frac{4}{3} f_a (\phi) E^a_{\phi} N^{-3} (E^3 -3 E E^{ij} E_{ij} + 2 E^{ij} E_{ik} E^k_{~j}) \biggr\} \biggr].
    \label{gbaction_adm_multi}	
\end{align}
where 
\begin{align}
    E_{ij} =&\,\, \frac{1}{2} \left(\dot{\gamma}_{ij} -\tilde{\nabla}_i N_j - \tilde{\nabla}_j N_i  \right), \quad E = \gamma^{ij} E_{ij},
    \\
    E_{\phi}^a =&\,\,  - \frac{1}{2} (\dot{\phi}^a - N^k \partial_k \phi^a), 	
\end{align}
$\tilde{\nabla}_i$ is the three dimensional covariant derivative with respect to  $\gamma_{ij}$, and $\tilde{R}_{ij}$ and $\tilde{R}$ are the Ricci tensor and Ricci scalarof the three-dimensional metric. The lapse $N$ and the shift $N^i$ appear without time derivatives and are therefore Lagrange multipliers enforcing the Hamiltonian and momentum constraints.

We work in the spatially flat gauge
\begin{align}
    \gamma_{ij} = a^2(t) \delta_{ij}, \,\, N = 1 + \alpha, \,\, N_i = \partial_i \psi, \,\, \delta \phi^a = Q^a.
\end{align}
Linearizing the action (\ref{gbaction_adm_multi}) in this gauge and expanding to the quadratic order yields
\begin{align}
    \mathcal{S}_s =&\,\, \int d^4x a^3 \biggl[ \mathcal{A} \alpha^2 - 2 \mathcal{F} \alpha a^{-2} (\partial^2 \psi) + \mathcal{B}_a D_t Q^a \alpha + \mathcal{G}_a Q^a \alpha - 8\lambda a^{-2} H^2 f_a (\partial^2 Q^a) \alpha
    \nonumber \\ 
    &\,\,   + \mathcal{K}_a a^{-2} Q^a  (\partial^2 \psi) + 8\lambda a^{-2} H^2 f_a D_t {Q}^a  (\partial^2 \psi) + \frac{1}{2} \bar{G}_{ab} (D_t Q^a) (D_t Q^b) + \frac{1}{2} \bar{R}_{acdb}\dot{\phi}_0^a \dot{\phi}_0^b  Q^c Q^d 
    \nonumber \\
    &\,\,  - \frac{1}{2} a^{-2} \delta^{ij} \bar{G}_{ab} \partial_i Q^a \partial_j Q^b  - \frac{1}{2} (D_a D_b V) Q^a Q^b + 12\lambda H^2 ( H^2 + \dot{H}) (D_a D_b f) Q^a Q^b
    \biggr],
   \label{quadsca-multib}	
\end{align}
where $\bar{R}_{abcd}$ is the Riemann tensor of the field space and the background coefficient functions are
\begin{align}
    \mathcal{F} =&\,\, 	M_{p}^2 H  + 12\lambda  H^2 f_a \dot{\phi}_0^a ,
    \\
    \mathcal{A} =&\,\, - 3 M_{p}^2 H^2 + \frac{1}{2} \bar{G}_{ab} \dot{\phi}_0^a \dot{\phi}_0^b - 48\lambda H^3 f_a \dot{\phi}_0^a,
    \\
    \mathcal{B}_a =&\,\, - \bar{G}_{ab} \dot{\phi}_0^b  + 24\lambda H^3 f_a,
    \\
    \mathcal{G}_a =&\,\, - V_a + 24\lambda H^3 (D_a D_b f) \dot{\phi}_0^b,
    \\
    \mathcal{K}_a =&\,\, \bar{G}_{ab} \dot{\phi}_0^b  - 8\lambda H^3 f_a + 8\lambda H^2 (D_a D_b f)\dot{\phi}_0^b,
\end{align}
and $D_a D_b f = f_{ab} - \Gamma^c_{ab} f_c$.

Varying with respect to $\alpha$ and $\partial^2 \psi$ gives the Hamiltonian and momentum constraint, 
\begin{align}
\mathcal{C}_0 =&\,\, a^3 \biggl[ 2 \mathcal{A} \alpha - 2 \mathcal{F} a^{-2} (\partial^2 \psi)  + \mathcal{B}_a D_t {Q}^a  - 8\lambda a^{-2} H^2 f_a (\partial^2 Q^a)  + \mathcal{G}_a Q^a  \biggr],
\label{ham-multia}
\\
C_1 =&\,\, a \biggl[ -2 \mathcal{F} \alpha + \mathcal{K}_a  Q^a  + 8\lambda H^2 f_a D_t Q^a  \biggr].
\label{mom-multia}
\end{align}
Solving the Hamiltonian constraint $\mathcal{C}_0 = 0$ from (\ref{ham-multia}) and the momentum constraint $\mathcal{C}_1 =0$ from (\ref{mom-multia}) for $(\partial^2 \psi)$ and $\alpha$ leads to
\begin{align}
	2 \mathcal{F} a^{-2} (\partial^2 \psi) =&\,\, 2 \mathcal{A} \alpha_1 + \mathcal{B}_a D_t Q^a - 8\lambda a^{-2} H^2 f_a (\partial^2 Q^a) + \mathcal{G}_a Q^a,
 \label{ham-multi}
 \\
    2 \mathcal{F} \alpha =&\,\, \mathcal{K}_a  Q^a + 8 \lambda H^2 f_a^{(0)} (D_t Q^a).
    \label{mom-multi}
\end{align}
Note that the momentum constraint determines the lapse perturbation algebraically in terms of $Q^a$ and $D_t Q^a$.  In particular, the Gauss-Bonnet coupling brings the field velocities directly into $\alpha$, which is the ultimate origin of the new velocity couplings found below.

Substituting the constraints (\ref{ham-multi}) and (\ref{mom-multi}) back into the quadratic action (\ref{quadsca-multib}), the quadratic action takes the form 
\begin{align}
     \mathcal{S}_s = \int d^4x a^3 \biggl[&\,\, \frac{1}{\mathcal{F}} \mathcal{D}_{ab} Q^a (D_t Q^b)  + \frac{1}{2} \mathcal{W}_{ab} (D_t Q^a) (D_t Q^b) -\frac{1}{2} \mathcal{Y}_{ab} a^{-2} (\partial_i Q^a) (\partial^i Q^b) 
    \nonumber \\
    &\,\,    -\frac{1}{2}  \biggl( (D_a D_b V) - 24 \lambda H^2 ( H^2 + \dot{H} ) (D_a D_b f)  - R_{dabc}\dot{\phi}_0^d \dot{\phi}_0^c 
    \nonumber \\
    &\,\, - \frac{1}{\mathcal{F}} \mathcal{K}_a \mathcal{G}_b  - \frac{\mathcal{A}}{2\mathcal{F}^2} \mathcal{K}_a \mathcal{K}_b \biggr) Q^a Q^b \biggr],
\end{align}
where
\begin{align}
	\mathcal{D}_{ab} =&	\frac{1}{2} \mathcal{K}_a \mathcal{B}_b + 4\lambda H^2 f_b \mathcal{G}_a + \frac{4\lambda \mathcal{A}}{\mathcal{F}} H^2  \mathcal{K}_a f_b,
\end{align}
which we decompose into its symmetric and antisymmetric parts, $\mathcal{D}_{ab} = \mathcal{D}_{(ab)} + \mathcal{D}_{[ab]}$, with $\mathcal{D}_{(ab)} =\frac{1}{2}  (\mathcal{D}_{ab} + \mathcal{D}_{ba})$ and $\mathcal{D}_{[ab]} = \frac{1}{2} (\mathcal{D}_{ab} - \mathcal{D}_{ba})$. The matrix $\mathcal{D}_{ab}$ is not symmetric, so the symmetric part can be integrated by parts into the mass matrix, while the antisymmetric part cannot be removed and survives as a genuine coupling. We, thus, arrive at  the final quadratic action for the field perturbations $Q^a$,
\begin{align}
     \mathcal{S}_s =&\,\, \int d^4x a^3 \biggl[\frac{1}{2} \mathcal{W}_{ab} (D_t Q^a) (D_t Q^b) + \mathcal{A}_{ab} Q^a (D_t Q^b) - \frac{1}{2} \mathcal{Y}_{ab} a^{-2} (\partial_i Q^a) (\partial^i Q^b) - \frac{1}{2} \mathcal{M}_{ab} Q^a Q^b 
    \biggr],
    \label{quadsca-multi}
\end{align}
where
\begin{align}
    \mathcal{A}_{ab} = \frac{1}{\mathcal{F}} \mathcal{D}_{[ab]},	
\end{align}
is antisymmetric, and  $\mathcal{W}_{ab},\,\, \mathcal{Y}_{ab}$ and $\mathcal{M}_{ab}$ are the symmetric kinetic, gradient, and mass matrices,
\begin{align}
	\mathcal{W}_{ab} =&\,\, \bar{G}_{ab} + \frac{8\lambda}{\mathcal{F}}  H^2 f_a \mathcal{B}_b + \frac{32\lambda^2}{\mathcal{F}^2} H^4 \mathcal{A} f_a  f_b,
    \label{wab}
\\
   \mathcal{Y}_{ab} =&\,\, \bar{G}_{ab} - \frac{8\lambda}{\mathcal{F}} H^2 f_a \mathcal{K}_b + \frac{32\lambda^2}{a} D_t \left(\frac{a}{\mathcal{F}} H^4 f_a f_b \right) 
   \label{yab}
   \\
   =&\,\, \mathcal{W}_{ab} + \frac{64\lambda^2}{\mathcal{F}^2} H^4 \biggl(- \dot{\phi}^2 + 12\lambda H^3 \dot{f} - 12\lambda H^2 \ddot{f} \biggr) f_a f_b,
\\
    \mathcal{M}^2_{ab} =&\,\, (D_a D_b V) - 24 \lambda H^2 \biggl( H^2 + \dot{H} \biggr) (D_a D_b f)  - \bar{R}_{dabc} \dot{\phi}_0^d \dot{\phi}_0^c 
    \nonumber \\
    &\,\, - \frac{1}{\mathcal{F}} \mathcal{K}_a \mathcal{G}_b - \frac{\mathcal{A}}{2\mathcal{F}^2} \mathcal{K}_a \mathcal{K}_b + \frac{1}{a^{3}}D_t\biggl(a^3\frac{1}{\mathcal{F}} \mathcal{D}_{(ab)} \biggr),
    \label{m2ab}
\end{align}
where symmetrization over the free indices $(ab)$ is understood in eqs. (\ref{wab}), (\ref{yab}), and (\ref{m2ab}) and the antisymmetric parts of the velocity couplings is contained entirely in $\mathcal{A}_{ab}$.

A comment on (\ref{quadsca-multi}) is in order. In the general relativistic (GR) limit $\lambda \rightarrow 0$, one has $\mathcal{K}_a \rightarrow \bar{G}_{ab} \dot{\phi}_0^b$ and $\mathcal{B}_a \rightarrow - \bar{G}_{ab} \dot{\phi}_0^b$, so that $\mathcal{D}_{ab} \rightarrow -\frac{1}{2} \dot{\phi}_{0a} \dot{\phi}_{0b}$ is symmetric and the antisymmetric coupling vanishes identically, $\mathcal{A}_{ab} \rightarrow 0$. The remaining matrices reduce to the standard multi-field expressions $\mathcal{W}_{ab},\, \mathcal{Y}_{ab} \rightarrow \bar{G}_{ab}$ with the familiar mass matrix \cite{GrootNibbelink:2001qt, Gong:2016qmq}. The antisymmetric velocity coupling $\mathcal{A}_{ab}$ is therefore a genuine Gauss-Bonnet effect. Unlike the antisymmetric couplings that can be generated in Einstein gravity by a time-dependent rotation of the field-space frame \cite{GrootNibbelink:2001qt, Lalak:2007vi,Achucarro:2010jv,Pinol:2020kvw}, it acts on the perturbations in the field space and cannot be removed by any such rotation. 
\subsection{Tensor perturbations}
Writing $\gamma_{ij} = a^2 (\delta_{ij} + h_{ij})$ with $h_{ij}$ transverse and traceless, the quadratic action for the tensor modes reads
\begin{align}
	\mathcal{S}_h = \int d^4x \frac{1}{8} a^3 &\,\, \mathcal{P} \biggl[ \dot{h}^{kl} \dot{h}_{kl}  - c_T^2 a^{-2} (\partial^k h^{ij}) (\partial_k h_{ij})  \biggr]
 \label{quadten-act}
\end{align}
where
\begin{align}
    c_T^2 = \frac{M_{p}^2 + 8\lambda\ddot{f}}{M_{p}^2 + 8\lambda H\dot{f}}, \quad \mathcal{P} = M_{p}^2 +8\lambda H\dot{f},
    \label{ct2}
\end{align}
and the second covariant time derivative of the coupling function along the trajectory is 
\begin{align}
    \ddot{f} = (D_a D_b f) \dot{\phi}_0^a \dot{\phi}_0^b + f_a D_t \dot{\phi}_0^a. 	
\end{align}
As in the single-field case, the Gauss-Bonnet coupling modifies both the normalization $\mathcal{P}$ and the propagation speed $c_T$ of the gravitational waves \cite{Satoh:2008ck, Guo:2010jr, Koh:2014bka, Bernardo:2025lie}. Stability of the tensor sector requires the absence of a ghost and of a gradient instability, $\mathcal{P} > 0$ and $c_T^2 \geq 0$. Both conditions  hold automatically in the weak-coupling regime, $\lambda \ll 1$, or in the slow-roll limits $|8\lambda H \dot{f}|,\, |8\lambda \ddot{f}| \ll M_{p}^2 $ considered below.  However, these stability conditions would impose significant constraints on strongly coupled models. 
\subsection{Curvature-isocurvature  mode decomposition}
To make contact with the observables, it is convenient to decompose the perturbations along and orthogonal to the background trajectory. For concreteness, we now specialize to two fields so that the isocurvature mode  is spanned by a single unit normal vector $N^a$. The generalization to more fields is straightforward. The field perturbations in flat gauge can be decomposed into the curvature perturbations $\mathcal{R}$ and the isocurvature perturbations $\mathcal{S}$ as \cite{Gordon:2000hv, Gong:2016qmq}
\begin{align}
    Q^a = N^a \mathcal{S} - \xi T^a \mathcal{R},	
    \label{decomp}
\end{align}
where the unit tangent vecotr $T^a$ is defined by
\begin{align}
    T^a = \frac{\dot{\phi}^a}{\dot{\phi}}, 
\end{align}
and $N^a$ is the unit normal, $\bar{G}_{ab} T^a N^b$ and $\bar{G}_{ab} N^a N^b =1$. For any field-space tensor $X_{ab}$, we write $X_{TT} = X_{ab} T^a T^b,\, X_{TN} = X_{ab}T^a N^b$ and $X_{NN} = X_{ab} N^a N^b$, and $\xi = \dot{\phi}_0/H$. Note that with the normalization (\ref{decomp}), the isocurvature perturbation $\mathcal{S}$ is a dimension of the fields and while $\mathcal{R}$ is dimensionless. 

Projecting $D_t T^a$ along $T^a$ and $N^a$ \cite{Gong:2016qmq}, we obtain
\begin{align}
    &\ddot{\phi} + 3H \dot{\phi} + V_T - 24\lambda H^2 (H^2 + \dot{H})f_T = 0,
    \\
    & D_t T^a = -\dot{\theta} N^a = \frac{1}{\dot{\phi}} \Big(- V_N + 24\lambda H^2 (H^2 + \dot{H}) f_N \Big) N^a,
\end{align}
so the trun rate of the trajectory is sourced by the normal gradient of the effective potential (\ref{eff_pot}), $\dot{\theta} = V_N^{\rm eff}/\dot{\phi}$ and the Gauss-Bonnet coupling bends the trajectory whenever $f_{N} \neq 0$. 

In terms of $\mathcal{R}$ and $\mathcal{S}$,  the quadratic action (\ref{quadsca-multi}) becomes
\begin{align}
    S_2 = \int d^4x \frac{a^3}{2} &\,\, \biggl[\mathcal{W}_{TT} \xi^2 \dot{\mathcal{R}}^2 -\mathcal{Y}_{TT} \xi^2 a^{-2} (\partial \mathcal{R})^2 - \xi^2 \mathcal{M}_{\mathcal{R}}^2 \mathcal{R}^2
    \nonumber \\
    &\,\, +\mathcal{W}_{NN}  \dot{\mathcal{S}}^2 - \mathcal{Y}_{NN} a^{-2} (\partial \mathcal{S})^2  - \mathcal{M}_{\mathcal{S}}^2 \mathcal{S}^2 
    \nonumber \\
    &\,\,   -  2 \mathcal{W}_{TN} \xi \dot{\mathcal{R}}\dot{\mathcal{S}}  + 2 \mathcal{Y}_{TN} \xi a^{-2} (\partial_i \mathcal{R}) (\partial^i \mathcal{S}) - 2 \xi \mathcal{M}_c^2 \dot{\mathcal{R}} \mathcal{S} -  2\xi \mathcal{M}^2_{\rm mix}  \mathcal{R} \mathcal{S} \biggr],
    \label{quadact_multi_ortho}
\end{align}
where
\begin{align}
    \mathcal{M}_{\mathcal{R}}^2 =&\,\, \frac{1}{a^3 \xi} D_t \Big( a^3 \mathcal{W}_{TT} \dot{\xi} \Big) - \frac{1}{a^3} D_t \Big( a^3 \mathcal{W}_{TN} \dot{\theta} \Big) - \mathcal{W}_{NN} \dot{\theta}^2 + \mathcal{M}^2_{TT} + 2 \mathcal{A}_{TN}\dot{\theta},
    \label{mr2-def}
    \\
    \mathcal{M}_{\mathcal{S}}^2 =&\,\, \frac{1}{a^3} D_t \Big( a^3  \mathcal{W}_{TN} \dot{\theta} \Big) -\mathcal{W}_{TT} \dot{\theta}^2  + \mathcal{M}^2_{NN} + 2\mathcal{A}_{TN} \dot{\theta} ,
    \label{ms2-def}
    \\
    \mathcal{M}_c^2 =&\,\, \mathcal{W}_{TT} \dot{\theta} - \mathcal{W}_{TN}\frac{\dot{\xi}}{\xi} +  \mathcal{W}_{NN} \dot{\theta} - 2\mathcal{A}_{TN},
    \label{mc2-def}
    \\
    \mathcal{M}^2_{\rm mix} =&\,\, \frac{\dot{\xi}}{\xi} \mathcal{W}_{TT} \dot{\theta} - \mathcal{W}_{TN} \dot{\theta}^2 - \frac{1}{a^3 \xi} D_t \big(a^3 \mathcal{W}_{TN} \dot{\xi} \big)  + \frac{1}{a^3 \xi} D_t \big(a^3 \xi \mathcal{W}_{NN} \dot{\theta} \big) 
    \nonumber \\
    &\,\, - \mathcal{M}_{TN}^2 - 2\frac{\dot{\xi}}{\xi} \mathcal{A}_{TN}- \frac{1}{a^3} D_t \left( a^3 \mathcal{A}_{TN}\right).
    \label{mmix2-def}
\end{align}

It is in order for discussing the stability conditions of the scalar modes. The absence of ghosts and gradient instabilities require both the kinetic and gradient matrix to be positive definite. In the orthogonal basis, one obtain the conditions
\begin{align}
    & \mathcal{W}_{TT} > 0,\quad \mathcal{W}_{TT} \mathcal{W}_{NN} - \mathcal{W}_{TN}^2 > 0,
    \nonumber \\ 
    & \mathcal{Y}_{TT} > 0, \quad \mathcal{Y}_{TT} \mathcal{Y}_{NN} - \mathcal{Y}_{TN}^2 > 0.
    \label{s_stable}
\end{align}
with the propagation speeds given by the eigenvalues of $\mathcal{W}^{-1} \mathcal{Y}$. We define the intrinsic sound speed of the curvature perturbation as
\begin{align}
    c_s^2 \equiv \frac{\mathcal{Y}_{TT}}{\mathcal{W}_{TT}}.
\end{align}
In the weak-coupling limit considered in the next section, the no-gohost condition  requires $\mathcal{W}_{TT} \simeq 1 - 8\lambda H \dot{f}/M_{p}^2  > 0$, which is satisfied within the perturbative regime $|8\lambda H\dot{f}/M_{p}^2|\ll 1$. Moreover, the difference between the gradient and kinetic coefficients starts at second order in $\lambda$,
\begin{align}
	\mathcal{Y}_{TT} - \mathcal{W}_{TT} = -64 \lambda^2 H^2 \dot{\phi}^2 f_T^2/M_{p}^4 + \mathcal{O}(\lambda^3).
\end{align}
Consequently, the propagation speed is 
\begin{align}
	c_s^2 =1 - 4\delta_1^2 + \mathcal{O}(\lambda^3),
\end{align}
where $\delta_1$ is a slow-roll parameter defined in the next section and is therefore slightly subliminal in the weak-coupling regime. At strong coupling, the conditions (\ref{s_stable}) becomes nontrivial and must be imposed on any concrete model.

Varying eq. (\ref{quadact_multi_ortho}) 
leads to the equations of motion of the curvature and isocurvature modes
\begin{align}
     \frac{d}{dt}\left(a^3 \mathcal{W}_{TT} \xi^2 \dot{\mathcal{R}}\right) & - a \mathcal{Y}_{TT}\xi^2 \partial^2 \mathcal{R} + a^3 \xi^2 \mathcal{M}_{\mathcal{R}}^2 \mathcal{R} 
    \nonumber \\
    & - \frac{d}{dt}\left(a^3 \mathcal{W}_{TN} \xi \dot{\mathcal{S}}\right) + a \mathcal{Y}_{TN} \xi \partial^2 \mathcal{S} - \frac{d}{dt}\left(a^3 \xi \mathcal{M}_c^2 \mathcal{S}\right) + a^3\xi \mathcal{M}_{\rm mix}^2 \mathcal{S} = 0,
    \label{eomr0}
    \\
     \frac{d}{dt}\left(a^3 \mathcal{W}_{NN} \dot{\mathcal{S}}\right) & - a \mathcal{Y}_{NN} \partial^2 \mathcal{S} + a^3 \mathcal{M}_S^2 \mathcal{S} 
    \nonumber \\
    & - \frac{d}{dt}\left(a^3 \mathcal{W}_{TN}\xi \dot{\mathcal{R}}\right) + a \mathcal{Y}_{TN}\xi \partial^2 \mathcal{R} + a^3\xi \mathcal{M}_c^2 \dot{\mathcal{R}} + a^3\xi \mathcal{M}_{\rm mix}^2 \mathcal{R} = 0.
    \label{eoms0}
\end{align}

In the GR limit ($\lambda \rightarrow 0)$, the coefficient function reduce to
\begin{align}
    &\mathcal{M}_{\mathcal{R}}^2  = 0, \quad \mathcal{M}_{\rm mix}^2 = 0,
    \\
    & \mathcal{M}_{\mathcal{S}}^2 = V_{NN} - \dot{\theta}^2 + \epsilon M_{p}^2 H^2 \mathbb{R},
    \\
    & \mathcal{M}_c^2 = 2\dot{\theta},
\end{align}
where $\epsilon = -\dot{H}/H^2$ and $\mathbb{R}$ is the Ricci scalar of the field space, and one recovers the standard two-field system \cite{Gordon:2000hv, Lalak:2007vi}. Remarkably, the first two statements are not restricted to the GR limit, but those hold identically for the full Gauss-Bonnet  theory once the background equations (\ref{friedmeq-multi}), (\ref{kgeq-multi}) , and (\ref{dH-multi}) are imposed. This is a natural consequence of the diffeomorphism invariance, which will be discussed in the next section.

\subsection{Weinberg's adiabatic mode}
The vanishing of $\mathcal{M}_{\mathcal{R}}^2$ and $\mathcal{M}_{\rm mix}^2$ is not an accidental feature of the Gauss-Bonnet theory. Rather, it follows from the diffeomorphism invariance of the theory and the existence of the adiabatic mode in the long-wavelength limit. The time-dependent rolling background $\phi^a(t)$ selects a preferred time slicing and thereby spontaneously breaks time diffeomorphism \cite{Cheung:2007st}. The corresponding Goldstone mode is the adiabatic perturbation along the background trajectory, which is represented by the comoving curvature perturbation $\mathcal R$, while the entropy perturbation $\mathcal S$ remains an independent physical degree of freedom. Weinberg's theorem \cite{Weinberg:2003sw} states that in the limit $k\rightarrow 0$ there always exists an adiabatic solution
\begin{align}
	(\mathcal{R}, \mathcal{S}) = (\text{const}., 0)
\end{align}
of the perturbation equations for any background.

If we impose $\mathcal{R} = \text{const}.$ and $\mathcal{S} = 0$ in eqs. (\ref{eomr0}) and (\ref{eoms0}) in the limit of $k\rightarrow 0$, one obtains 
\begin{align}
    	a^3 \xi^2 \mathcal{M}_{\mathcal{R}}^2 = 0, \quad \xi \mathcal{M}_{\rm mix}^2 \mathcal{R} = 0,
\end{align}
and since these must hold on an arbitrary background,  we conclude that
\begin{align}
    \mathcal{M}_{\mathcal{R}}^2 = 0, \quad \mathcal{M}_{\rm mix}^2 = 0.	
    \label{ward}
\end{align}
The vanishing of these two masses provides a stringent check on the long and intricate computation leading to eqs. (\ref{mr2-def}) -- (\ref{mmix2-def}). 

Using eq. (\ref{ward}), the quadratic action (\ref{quadact_multi_ortho}) takes the final form
\begin{align}
    S_2 = \int d^4x \frac{a^3}{2} &\,\, \biggl[\mathcal{W}_{TT} \xi^2 \dot{\mathcal{R}}^2 -\mathcal{Y}_{TT} \xi^2 a^{-2} (\partial \mathcal{R})^2 +\mathcal{W}_{NN}  \dot{\mathcal{S}}^2 - \mathcal{Y}_{NN} a^{-2} (\partial \mathcal{S})^2  - \mathcal{M}_{\mathcal{S}}^2 \mathcal{S}^2 
    \nonumber \\
    &\,\,   -  2 \mathcal{W}_{TN} \xi \dot{\mathcal{R}}\dot{\mathcal{S}}  + 2 \mathcal{Y}_{TN} \xi a^{-2} (\partial_i \mathcal{R}) (\partial^i \mathcal{S}) - 2 \xi \mathcal{M}_c^2 \dot{\mathcal{R}} \mathcal{S}\biggr],
    \label{quadact_multi_ortho_fin}
\end{align}
and correspondingly, the  equations of motion for the curvature and isocurvature modes become 
\begin{align}
     \frac{d}{dt}\left(a^3 \mathcal{W}_{TT} \xi^2 \dot{\mathcal{R}}\right) & - a \mathcal{Y}_{TT}\xi^2 \partial^2 \mathcal{R}
    \nonumber \\
    & - \frac{d}{dt}\left(a^3 \mathcal{W}_{TN} \xi \dot{\mathcal{S}}\right) + a \mathcal{Y}_{TN} \xi \partial^2 \mathcal{S} - \frac{d}{dt}\left(a^3 \xi \mathcal{M}_c^2 \mathcal{S}\right)  = 0,
    \label{eomr}
    \\
     \frac{d}{dt}\left(a^3 \mathcal{W}_{NN} \dot{\mathcal{S}}\right) & - a \mathcal{Y}_{NN} \partial^2 \mathcal{S} + a^3 \mathcal{M}_S^2 \mathcal{S} 
    \nonumber \\
    & - \frac{d}{dt}\left(a^3 \mathcal{W}_{TN}\xi \dot{\mathcal{R}}\right) + a \mathcal{Y}_{TN}\xi \partial^2 \mathcal{R} + a^3\xi \mathcal{M}_c^2 \dot{\mathcal{R}} = 0.
    \label{eoms}
\end{align}
In particular, eq. (\ref{eomr}) in the super-horizon scale limit ($k\rightarrow 0$) becomes a conservation law
\begin{align}
	 \frac{d}{dt}\left[a^3 \left(\mathcal{W}_{TT} \xi^2 \dot{\mathcal{R}}-  \mathcal{W}_{TN} \xi \dot{\mathcal{S}}- \xi \mathcal{M}_c^2 \mathcal{S}\right) \right]  = 0,
\end{align}
or, dropping the decaying solution 
\begin{align}
    \dot{\mathcal{R}} = \frac{\mathcal{W}_{TN}}{\mathcal{W}_{TT}\xi} \dot{\mathcal{S}} + \frac{\mathcal{M}_c^2}{\mathcal{W}_{TT}\xi} \mathcal{S}.	\label{largek_consrve}
\end{align}
The curvature perturbation on large scale is therefore constant unless it is sourced by the isocurvature mode, exactly as in Einstein gravity, but with the source corrected by the Gauss-Bonnet terms, $\mathcal{W}_{TN},\, \mathcal{W}_{TT},$ and $\mathcal{M}_c^2$.
This reduces in the GR limit ($\lambda \rightarrow 0$) \cite{Gordon:2000hv}
\begin{align}
    \dot{\mathcal{R}} = \frac{2\dot{\theta}}{\xi} \mathcal{S},	
\end{align}
where we have used $\mathcal{W}_{TT} = 1,\, \mathcal{W}_{TN} =  0, $ and $\mathcal{M}_c^2 = 2\dot{\theta}$ in the GR limit.

\section{Power spectra} \label{sect:spectra}

The quadratic action (\ref{quadact_multi_ortho_fin}) keeps the kinetic coupling, the spatial-gradient coupling, the velocity coupling and the mixing mass terms. These terms make it difficult to analyze the dynamics, obtain the observables, and  interpret them in full generality. In this section, we therefore consider two physically motivated limiting cases and calculate the power spectrum and the observables. Throughout this section, we use the slow-roll conditions and define the slow-roll parameters as
\begin{align}
    \epsilon =&\,\, - \frac{\dot{H}}{H^2}, \quad \eta_{\parallel} = \frac{\ddot{\phi}}{H \dot{\phi}}, \quad \eta_{\perp} = \frac{\dot{\theta}}{H}, \quad \delta_1 = \frac{4\lambda H\dot{f}}{M_{p}^2} = \frac{4\lambda H\dot{\phi} f_T}{M_{p}^2}, 
    \nonumber \\
    \delta_{N} =&\,\, \frac{4\lambda H \dot{\phi} f_N}{M_{p}^2 }, \quad \delta_{TT} = 8\lambda H^2 f_{TT}, \quad \delta_{TN} = 8\lambda H^2 f_{TN}, \quad \delta_{NN} = 8\lambda H^2 f_{NN},
    \label{srparams}
\end{align}
where $f_{TT},\,f_{TN}$ and $f_{NN}$ are the projections of $D_a D_b f$. 

\subsection{Heavy isocurvature mode}
When the isocurvature mode is heavy,  $\mathcal{M}_{\mathcal{S}} \gg H^2 $, it can be integrated out on all scales\cite{Achucarro:2010da, Cespedes:2012hu}. Neglecting the temporal and spatial derivative terms of $\mathcal{S}$ in eq. (\ref{eoms}), together with the derivative mixings $\mathcal{W}_{TN}$ and $\mathcal{Y}_{TN}$, which are of order $\delta_N$ and slow-roll suppressed,  we get the relation
\begin{align}
	\mathcal{S} \simeq -\xi \frac{\mathcal{M}_c^2}{\mathcal{M}_{\mathcal{S}}^2} \dot{\mathcal{R}}.
\end{align}
Eliminating $\mathcal{S}$ from the quadratic action (\ref{quadact_multi_ortho_fin}), ignoring the temporal and spatial derivatives of the isocurvature mode, leaves the effective  single field action
\begin{align}
    S_{\rm eff} = \int dt d^3x \frac{a^3}{2} \xi^2 \Sigma \left[ \dot{\mathcal{R}}^2 - \frac{c_{s,{\rm eff}}^2}{a^2} (\partial \mathcal{R})^2\right],	
    \label{seff}
\end{align}
where
\begin{align}
    \Sigma = \mathcal{W}_{TT} + \frac{\mathcal{M}_c^4}{\mathcal{M}_s^2},\quad  \frac{1}{c_{s, {\rm eff}}^2} = \frac{1}{c_s^2} \left[1+\frac{\mathcal{M}_{c}^4}{\mathcal{W}_{TT} \mathcal{M}_{\mathcal{S}}^2}\right],
    \label{cseff}
\end{align}
and $c_s^2 = \mathcal{Y}_{TT}/\mathcal{W}_{TT}$ is the intrinsic adiabatic sound speed. As in Einstein gravity, integrating out the heavy isocurvature mode reduces the effective sound speed, $c_{s,{\rm eff}} \leq c_s$.

The action (\ref{seff}) is quantized  in the standard way in terms of the canonical variable
\begin{align}
    v = z \mathcal{R}, \quad z = a\xi \sqrt{\Sigma} = \frac{a \dot{\phi}}{H}\sqrt{\mathcal{W}_{TT} + \frac{\mathcal{M}_c^4}{\mathcal{M}_{\mathcal{S}}^2}}.	
\end{align}
Following the standard procedure with the Bunch-Davies initial vacuum state, the positive frequency mode functions are
\begin{align}
    v_k(\tau) = \frac{e^{-i c_{s,{\rm eff}} k\tau}}{\sqrt{2 c_{s,{\rm eff}} k}} \left(1-\frac{i}{c_{s,{\rm eff}}k\tau}\right), 	
\end{align}
where $\tau$ is the conformal time, and the curvature power spectrum evaluated at sound-horizon exit is
\begin{align}
    P_{\mathcal{R}}(k) = \frac{k^3}{2\pi^2} \left|\frac{v_k}{z} \right|^2	= \left. \frac{H^4}{4\pi^2 \dot{\phi}^2 \Sigma c_{s,{\rm eff}}^3} \right|_{c_{s,{\rm eff}} k = aH} = P_{\mathcal{R}}^* \frac{c_s}{c_{s,{\rm eff}}} = P_{\mathcal{R}}^* \sqrt{1+\frac{\mathcal{M}_c^4}{\mathcal{W}_{TT} \mathcal{M}_{\mathcal{S}}^2}}	
    \label{prheavy}
\end{align}
where 
\begin{align}
    P_{\mathcal{R}}^* = \frac{H^4}{4\pi^2 \dot{\phi}^2 \mathcal{W}_{TT} c_s^3 }	
\end{align}
is the spectrum that would be obtained in the absence of the coupling to the heavy mode. The heavy isocurvature mode therefore always enhances the curvature spectrum by the factor $c_s/c_{s,{\rm eff}}$.

Since $\Sigma c_{s,{\rm eff}}^3 = \mathcal{W}_{TT} c_s^2 c_{s,{\rm eff}}$ by eq. (\ref{cseff}), the spectrum (\ref{prheavy}) can be written as
\begin{align}
    P_{\mathcal{R}} (k) = \frac{H^4}{4\pi^2 \dot{\phi}^2 \mathcal{W}_{TT} c_s^2 c_{s,{\rm eff}}}.
    \label{sspectra}
\end{align}
Taking the logarithmic derivative at sound-horizon exit, where $d\ln k \simeq d N$ at leading order in slow-roll, using $c_s^2 = 1 + \mathcal{O}(\lambda^2)$, the scalar spectral index at linear order in $\lambda$ in the weak-coupling limit ($\lambda \ll 1$) is
\begin{align}
    n_{\mathcal{R}}-1 = -4\varepsilon - 2\eta_{\parallel} + 2\left(\delta_1 (\eta_{\parallel}-\varepsilon) +\varepsilon \delta_{TT} - \frac{\dot{\theta}}{H} \delta_{N} \right) - \frac{d}{dN} \ln c_{s,{\rm eff}} + \mathcal{O}(\lambda^2),
\end{align}
where we have used
\begin{align}
    \frac{d \delta_1}{dN} = \delta_1 (\eta_{\parallel} - \epsilon) + \epsilon \delta_{TT} - \frac{\dot{\theta}}{H}\delta_N,
    \label{dd1dn}
\end{align}
and $\dot{f}_T = \dot{\phi} f_{TT} - \dot{\theta} f_N$. The first two terms are the standard single-field tilt, the round bracket is the tilt by the Guass-Bonnet parameter along the trajectory, and the last term is due to the effective sound speed.

Finally, in the heavy-isocurvature regime, the entropy mode oscillates and dilutes after horizon exit, with $\mathcal{S} \propto a^{-3 / 2}$ and hence $\mathcal{P}_{\mathcal{S}} \propto a^{-3}$ \cite{Gordon:2000hv, Geller:2022nkr}. Thus, if the mode remains heavy until the end of inflation, the isocurvature perturbation is negligibly small.

\subsection{$f_N =0$ and weak coupling}

The second limit we consider is the case in which the gradient of the coupling function has no component normal to the background trajectory, $f_N = 0$. Because  $\mathcal{B}_N = 24\lambda H^3 f_N = 0$, imposing $f_N=0$ gives  $W_{TN} = Y_{TN} = 0$ and $W_{NN} = Y_{NN} = 1$ as well as $c_s^2 = 1$. All derivative mixing between curvature and isocurvature modes disappear.

With $f_N=0$, the curvature (\ref{eomr}) and isocurvature equations (\ref{eoms}) become
\begin{align}
     \frac{d}{dt}\left(a^3 \mathcal{W}_{TT} \xi^2 \dot{\mathcal{R}}\right) & - a \mathcal{Y}_{TT}\xi^2 \partial^2 \mathcal{R} =  \frac{d}{dt}\left(a^3 \xi \mathcal{M}_c^2 \mathcal{S}\right) 
    \label{eomr-fn0}
    \\
     \frac{d}{dt}\left(a^3 \mathcal{W}_{NN} \dot{\mathcal{S}}\right) & - a \mathcal{Y}_{NN} \partial^2 \mathcal{S} + a^3 \mathcal{M}_S^2 \mathcal{S} = - a^3\xi \mathcal{M}_c^2 \dot{\mathcal{R}}.
    \label{eoms-fn0}
\end{align}
where
\begin{align}
    \mathcal{M}_c^2 =&\,\, (\mathcal{W}_{TT} + 1) \dot{\theta} -2 A_{TN},
    \\
    \mathcal{M}_s^2 =& \mathcal{M}_{NN}^2 - \mathcal{W}_{TT} \dot{\theta}^2 + 2A_{TN} \dot{\theta} .
\end{align}

On superhorizon scales, eqs. (\ref{eomr-fn0}) and (\ref{eoms-fn0}) reduce to first order equation in terms of the number of $e$-folds neglecting the spagial gardient terms. From eq. (\ref{largek_consrve}) with $\mathcal{W}_{TN} = 0$,
\begin{align}
    \frac{d\mathcal{R}}{dN} = \alpha (N) \mathcal{S}, \quad \frac{d \mathcal{S}}{dN} = \beta(N) \mathcal{S}, \quad \alpha(N) = \frac{\mathcal{M}_c^2}{H\xi \mathcal{W}_{TT}},	
\end{align}
where $\beta$ follows from the large-scale limit of eq. (\ref{eoms-fn0}), by neglecting $\ddot{\mathcal{S}}$ due to the slow-roll conditions and eliminating $\dot{\mathcal{R}}$ with eq. (\ref{largek_consrve}),
\begin{align}
    \beta(N) \simeq -\frac{1}{3H^2}\left(\mathcal{M}_{\mathcal{S}}^2 + \frac{\mathcal{M}_c^4}{\mathcal{W}_{TT}}\right).
    \label{betan}
\end{align}

If the transfer matrix is defined on superhorizon scales by \cite{Gordon:2000hv,Wands:2000dp,Byrnes:2006fr, Peterson:2011yt,Karamitsos:2017elm}
\begin{align}
    \left(\begin{array}{c} \mathcal{R} \\ \mathcal{S}	\end{array}\right)_{end} = \left(\begin{array}{lr} 1 & T_{\mathcal{R} \mathcal{S}} \\ 0 & T_{\mathcal{S} \mathcal{S}} \end{array} \right) \left(\begin{array}{c} \mathcal{R} \\ \mathcal{S}	\end{array}\right)_{*},
\end{align}
where the asterisk denotes the horizon crossing, and assuming the horizon-crossing modes are initially uncorrelated $\langle \mathcal{R}_{*} \mathcal{S}_{*}\rangle = 0$, the curvature and isocurvature power spectra at the end of inflation  become
\begin{align}
    P_{\mathcal{R}} =&\,\, P_{\mathcal{R}_*} + T^2_{\mathcal{R} \mathcal{S}} P_{\mathcal{S}_*},	
    \\ 
    P_{\mathcal{S}} =&\,\, T^2_{\mathcal{S} \mathcal{S}} P_{\mathcal{S}_*},	
\end{align}
where the transfer functions $T_{\mathcal{R} \mathcal{S}}$ and $T_{\mathcal{S} \mathcal{S}}$ are given by
\begin{align}
    T_{\mathcal{R} \mathcal{S}} (N, N_*) =&\,\,  \int^N_{N_*} dN' \alpha (N') T_{\mathcal{S} \mathcal{S}}(N', N_*),
\\
    T_{\mathcal{S} \mathcal{S}} (N, N_*) =&\,\, \exp \left[\int^N_{N_*} \beta (N') dN' \right].	
\end{align}

The power spectrum at the end of inflation is calculated explicitly for $\lambda \ll 1$. The coefficient functions for general $f_N$ are, at linear order in $\lambda$ and in terms of the slow-roll parameters (\ref{srparams}),
\begin{align}
    \mathcal{W}_{TT} =&\,\, \mathcal{Y}_{TT} = 1 - 2 \delta_1, \quad \mathcal{W}_{NN} = \mathcal{Y}_{NN} = 1,
    \quad \mathcal{W}_{TN} = \mathcal{Y}_{TN} = -\delta_N,
    \\
    A_{TN} =&\,\, \frac{1}{2} \dot{\theta} \delta_1 + H\delta_N  + \frac{1}{2} H (\epsilon+\eta_{\parallel})\delta_N + \frac{1}{2} H \epsilon \delta_{TN}, 
    \\
    \mathcal{M}_c^2 =&\,\, 2\dot{\theta} -3 \dot{\theta}\delta_1 -2H \delta_N -H\epsilon \delta_{TN},
    \\
    \mathcal{M}_{\mathcal{S}}^2 =&\,\, V_{NN} - \dot{\theta}^2 + \epsilon M_{p}^2 H^2 \mathbb{R} - 3H^2 (1-\epsilon)\delta_{NN} -10H \dot{\theta} \delta_N + 2\dot{\theta}^2 \delta_1
    \nonumber \\
    & - 2\ddot{\theta} \delta_N + 2H\dot{\theta}(\epsilon - \eta_{\parallel})\delta_N + 2H \dot{\theta} \epsilon \delta_{TN}. \label{ms2-sr}
\end{align}
Note, in particular, the term $-3H^2 (1-\epsilon) \delta_{NN}$ in eq. (\ref{ms2-sr}).  The normal-normal component of the coupling function shifts the isocurvature mass at leading order in slow-roll  and can make the isocurvature direction heavier or lighter (or even tachyonic) depending on the sign of $f_{NN}$, even when $f_N = 0$.

With $\delta_N = 0$, keeping only first-order slow-roll terms, we have
\begin{align}
    \mathcal{M}_c^2 \simeq 2\dot{\theta},	
\end{align}
and therefore
\begin{align}
    \alpha \simeq \frac{2\dot{\theta}}{H\xi (1-2\delta_1)} = \frac{2\eta_{\perp}}{\xi (1-2\delta_1)},	
\end{align}
while eq. (\ref{betan}) becomes
\begin{align}
    \beta \simeq -\frac{V_{NN}}{3H^2} - \eta_{\perp}^2 - \frac{\epsilon}{3} M_{p}^2 \mathbb{R} + (1-\epsilon) \delta_{NN} + \frac{2}{3} \eta_{\perp}^2 \delta_1 + \frac{2}{3} \epsilon \eta_{\perp} \delta_{TN}.	
\end{align}
The leading Gauss-Bonnet correction to the superhorizon decay of the isocurvature mode is the $\delta_{NN}$ term. For $\lambda f_{NN} > 0$, it slows the decay of $T_{\mathcal{S} \mathcal{S}}$, and can even destabilize the isocurvature direction, enhancing the transfer computed below, while $\lambda f_{NN} < 0$ suppresses it. The curvature power spectrum  at horizon crossing, to first order in slow-roll parameters, is 
\begin{align}
    P_{\mathcal{R}_*} \simeq \frac{H_*^2}{8\pi^2 M_{p}^2 \epsilon_*} (1+2\delta_1).	
\end{align}
Similarly, for a light isocurvature mode, $|\mathcal{M}_{\mathcal{S}}^2|\ll H^2$, eq. (\ref{eoms-fn0}) with $\mathcal{W}_{NN} = \mathcal{Y}_{NN} = 1$ describes a canonically normalized, effectively massless spectator around horizon crossing, so that
\begin{align}
    P_{\mathcal{S}_*}(k) \simeq \left(\frac{H_*}{2\pi}\right)^2 \left[ 1 + \mathcal{O}(\epsilon, \mathcal{M}_{\mathcal{S}}^2/H^2, \delta) \right].
\end{align}
Here, the $\mathcal{M}_c^2$ mixing is $\mathcal{O}(\eta_{\perp})$ which contributes to the cross-correlation and contributes the auto-correlation only at $\mathcal{O}(\eta_{\perp}^2)$. Thus, at the end of inflation, 
\begin{align}
   P_\mathcal{R}^{end}(k) \simeq =  	 \frac{H_*^2}{8\pi^2 M_{p}^2 \epsilon_*} (1+2\delta_1) + T_{\mathcal{R} \mathcal{S}}^2 (k) P_{\mathcal{S}_*} (k),
    \label{prdn0}
\end{align}
with
\begin{align}
    T_{\mathcal{R} \mathcal{S}} = \int^{N_{end}}_{N_*} dN \frac{\eta_{\perp}}{(1-2\delta_1)\xi} T_{\mathcal{S} \mathcal{S}}(N, N_*).
    \label{transfer_trs}
\end{align}
The transfer of isocurvature mode into curvature perturbations is thus governed  by the turn rate $\eta_{\perp}$, ehnhaced by the Gauss-Bonnet factor $(1-2\delta_1)^{-1}$, while $T_{\mathcal{S} \mathcal{S}}$ is controlled by the isocurvature mass (\ref{ms2-sr}). 

\subsection{Tensor spectrum}
For completeness we calculate the observables of the tensor modes. Quantizing the action (\ref{quadten-act}) in the standard way, the tensor power spectrum evaluated at horizon-crossing is
\begin{align}
   P_T(k) =  \left. \frac{2H^2}{\pi^2 \mathcal{P} c_T^3} \right|_{c_T k = aH}.
\end{align}
At linear order in $\lambda$, using $\mathcal{P} = M_{p}^2 (1+2\delta_1)$ and $\ddot{f} = \dot{\phi}^2 f_{TT} + \ddot{\phi} f_T - \dot{\phi} \dot{\theta} f_N$, the tensor propagation speed (\ref{ct2}) to leading order in slow-roll read as
\begin{align}
    c_T^2 = 1 - 2\delta_1.	
\end{align}
Then the tensor spectrum becomes
\begin{align}
    P_T(k) \simeq \frac{2H_*^2}{\pi^2 M_{p}^2} (1+\delta_1),	
    \label{tspectra}
\end{align}
and the tensor spectral index is calculated as
\begin{align}
    n_T \simeq -2\epsilon + \frac{d \delta_1}{dN}, 
\end{align}
where $d\delta_1/dN$ is given in eq. (\ref{dd1dn}). Combining eq. (\ref{tspectra}) with the scalar spectra of the two limits gives the tensor-to-scalar ratio. In the heave isocurvature regime, eq. (\ref{sspectra}) with $c_s^2 =1$ yields
\begin{align}
    r = \frac{P_T(k)}{P_{\mathcal{R}}(k)} \simeq 16 \epsilon c_{s,{\rm eff}} (1-\delta_1),	
\end{align}
generalizing the familiar $r=16\epsilon c_s$ suppression of heavy-field effective theories \cite{Achucarro:2010da} by the Gauss-Bonnet corrections. In the $f_N=0$ limit with the light isocurvature mode, eq. (\ref{prdn0}) gives
\begin{align}
    r \simeq \frac{16 \epsilon_* (1-\delta_{1*})}{1+ T_{\mathcal{R} \mathcal{S}}^2 P_{\mathcal{S}_*}/P_{\mathcal{R}_*}},	
\end{align}
so the superhorizon transfer suppresses $r$.

\section{Summary and Discussions} \label{sect:summary}
In this paper we have studied curvature and isocurvature perturbations in multi-field inflation with a Gauss-Bonnet coupling. Considering a nonlinear sigma model on a curved field space coupled to the Gauss–Bonnet invariant through an arbitrary function $f(\phi^a)$, we derived the complete quadratic action for the field perturbations $Q^a$ in the spatially flat gauge using the ADM formalism and solving the Hamiltonian and momentum constraints. In addition to the kinetic, gradient, and mass terms, the quadratic action contains a velocity coupling of the form $\mathcal{D}_{ab} Q^a D_t Q^b$ , whose coupling matrix is generally not symmetric. In particular, the Gauss–Bonnet interaction generates a nonvanishing antisymmetric component $\mathcal{D}_{[ab]}$, originating from the modified constraint equations and their dependence on the background field velocities. This antisymmetric coupling vanishes identically in the Einstein limit and cannot be eliminated by an orthogonal rotation of the field-space frame. It therefore represents a genuine dynamical effect of the Gauss–Bonnet interaction and provides a new channel for curvature–isocurvature mixing in multi-field inflation.

After decomposing the field perturbations into curvature and isocurvature modes in an orthonormal basis, we find that the effective mass of the curvature perturbation and  the curvature–isocurvature mass mixing vanish identically. This cancellation is a consequence of the residual diffeomorphism symmetry associated with the Weinberg adiabatic mode in the $k\rightarrow 0$ limit. Consequently, the superhorizon dynamics is governed by the conservation law (\ref{largek_consrve}), according to which the curvature perturbation remains conserved on large scales in the absence of isocurvature sourcing. The Gauss–Bonnet coupling does not spoil this adiabatic conservation law, but instead modifies the transfer from the isocurvature mode to the curvature perturbation through the GB-corrected source term.

We then computed the observables in two limits. First, when the entropic mode is heavy, the curvature perturbation is described by the effective single field action with a reduced effective sound speed. The resulting spectrum is enhanced by the factor $c_s/c_{s,eff}$ and the spectral index receives Gauss-Bonnet corrections through the slow-roll parameters $\delta_1,\,\delta_{TT},$ and $\delta_N$. Second, when the gradient of the coupling function is aligned with the trajectory, $f_N = 0$,  all derivative mixings between the curvature and isocurvature modes vanish, and the superhorizon evolution of the curvature perturbation is determined by the transfer function $T_{\mathcal{R} \mathcal{S}}$ (\ref{transfer_trs}) which is governed by the turn rate $\eta_{\perp}$ and enhanced by the Gauss-Bonnet factor $(1-2\delta_1)^{-1}$. The isocurvature curvature spectrum is determined by $T_{\mathcal{S} \mathcal{S}}$ which is controlled by the mass of the isocurvature mode. In both limiting cases, we have presented the tensor spectrum and the tensor-to-scalar ratio at linear order in $\lambda$ and established the corresponding no-ghost and gradient-stability conditions.

Several directions follow from this work.  First, the enhancement of the curvature spectrum through the reduced effective sound speed in the heavy isocurvature mode regime suggests that multi-field Gauss-Bonnet inflation may provide a mechanism for enhancing the small-scale scalar power spectrum relevant for primordial black hole formation and the associated scalar-induced gravitational waves. A quantitative investigation of the resulting small-scale spectrum and its phenomenological constraints is an interesting direction for future work.  Second, the effects of the Gauss-Bonnet coupling on primordial non-Gaussianity remain to be explored. In particular, the antisymmetric coupling $\mathcal{A}_{ab}$ and the velocity coupling $\mathcal{M}_c^2$ may generate characteristic modifications to the interactions between curvature and isocurvature modes, with possible implications for quasi-single-field inflation and cosmological-collider signals. Finally, it would be interesting to study the rapid-turn regime, in which  the couplings derived here are no longer perturbatively small and the interplay between the Gauss-Bonnet sector, field-space geometry, and trajectory turning may lead to qualitatively new dynamics.


\acknowledgments

SK would like to thank T.~Tanaka for his valuable comments  while the author was attending the JGRG34 workshop, and G.~Tumurtushaa and M.~Mishra for their useful discussions. This work is supported by the National Research Foundation of Korea (NRF) under grants NRF-2021R1A2C1005748.



\bibliographystyle{JHEP}
\bibliography{refs}

\end{document}